\input{aipcheck}

\documentclass[
    ,final            
  ]
  {aipproc}

\layoutstyle{6x9}

\begin{document}

\title{Status and Prospects of {\it Fermi} LAT Pulsar Blind Searches}


\author{P.~M.~Saz Parkinson for the $Fermi$-LAT Collaboration\footnote{\tt{http://www-glast.stanford.edu/cgi-bin/people}}}{
  address={Santa Cruz Institute for Particle Physics, University of California, Santa Cruz, CA 95064}
  ,altaddress={e-mail: pablo@scipp.ucsc.edu}
}

\begin{abstract}
Blind Searches of {\it Fermi} Large Area Telescope (LAT) data have resulted in the discovery of 24 $\gamma$-ray 
pulsars in the first year of survey operations, most of which remain undetected in radio, despite deep radio follow-up 
searches. I summarize the latest {\it Fermi} LAT blind search efforts and results, including the discovery of a new Geminga-like 
pulsar, PSR~J0734--1559. Finally, I discuss some of the challenges faced in carrying out these searches into the future, 
as well as the prospects for finding additional pulsars among the large number of LAT unassociated sources.
\end{abstract}

\maketitle


{\bf Introduction -- } The {\it Fermi} Large Area Telescope (LAT), launched on 11 June 2008, represents a giant leap in capabilities, relative to 
past $\gamma$-ray missions. Its larger field of view ($\sim$2 sr), effective area (9500 cm$^2$, @1 GeV, normal incidence), better point spread function 
($\sim0.6^\circ$ @1 GeV), broader energy range (20 MeV -- 300 GeV), and significantly shorter deadtime, result in a vastly 
improved sensitivity. In addition, the LAT operates in continuous sky survey mode, efficiently scanning the sky with very uniform coverage. These 
improvements have led to the discovery of a large number of $\gamma$-ray sources, while also greatly improving the 
characterization of $\gamma$-ray sources detected by past missions (e.g. EGRET). For a detailed description of the LAT, see \cite{Fermi}.

{\bf Blind Searches of LAT data -- } The improved sensitivity of the LAT has resulted in the detection of an order of magnitude more $\gamma$-ray pulsars 
than were previously known (see Romani, these proceedings). In addition to detecting $\gamma$-ray pulsations from known radio pulsars (e.g. Guillemot et al., 
Parent et al., Belfiore, in these proceddings), the LAT is the 
first $\gamma$-ray telescope to independently discover pulsars through blind searches of $\gamma$-ray data. Searching for pulsars in 
$\gamma$-ray data poses significant challenges, the main one being a scarcity of photons. Despite its huge improvement in sensitivity, the LAT still 
only detects a relatively small number of $\gamma$-ray photons from a given source. For example, the LAT 
detects fewer than 1000 ($>$30 MeV) photons per day (fewer than 1 per 1,000 rotations) from the Vela pulsar, the brightest steady $\gamma$-ray 
source in the sky~\citep{LATVela}. Typical $\gamma$-ray pulsars result in tens or at most hundreds of 
photons per day. The detection of $\gamma$-ray pulsations therefore requires observations spanning long periods of time (up to years), during 
which the pulsars not only slow down, but often also experience significant timing irregularities, such as "timing noise" or glitches.

{\bf The time differencing technique} -- 
In order to lessen the impact of the long integrations required for blind searches of $\gamma$-ray pulsars, a new technique, known as 
"time-differencing"~\cite{2006ApJ...652L..49A,2008ApJ...680..620Z}, was developed, in which FFTs are computed on the time differences of events, 
rather than the times themselves. By limiting the maximum time window up to which differences are computed to $\sim$days, rather than 
months or years, the required number of FFT bins is greatly reduced. The reduced frequency resolution results in a larger step 
size required in frequency derivative, $\dot{f}$, thus greatly reducing the number of $\dot{f}$ trials needed to cover the requisite parameter space, with 
the added bonus of making such searches less sensitive to timing irregularities than a traditional coherent search. The net result is 
a significant reduction in the computational and memory costs, relative to the standard FFT methods, with only a modest effect on the overall 
sensitivity~\cite{2006ApJ...652L..49A}. 

{\bf The first 24 blind search pulsars} -- Within a few months of the launch of $Fermi$, sixteen $\gamma$-ray pulsars had been discovered using the 
time-differencing technique on LAT data~\citep{2009Sci...325..840A}. Subsequently, an additional 
8 pulsars were discovered in blind searches of $\sim$1 year of LAT data~\citep{2010ApJ...725..571S}, bringing the total number of pulsars 
discovered by the LAT to 24 in its first year of survey operations. A large number of these pulsars were found to be coincident with previously 
unidentified EGRET sources, thereby leading to their identification. Many of these pulsars are young and energetic, and are associated 
with pulsar wind nebulae (PWNe) or supernova remnants (SNRs), which in many cases are detected in other wavelengths, such as X-rays or 
TeV energies (e.g. \cite{milagro}). Deep radio observations of all 24 of these pulsars resulted in the detection of only three of them, with one of 
the three pulsars (PSR J1907+0602) showing up as extremely faint in radio (flux density of $\sim3.4\mu$Jy), raising the possibility that other 
pulsars considered radio quiet may, in fact, be emitting in radio, only at much lower fluxes than hitherto expected~\cite{J1907}.

\begin{figure}[h]
  \includegraphics[height=.25\textheight]{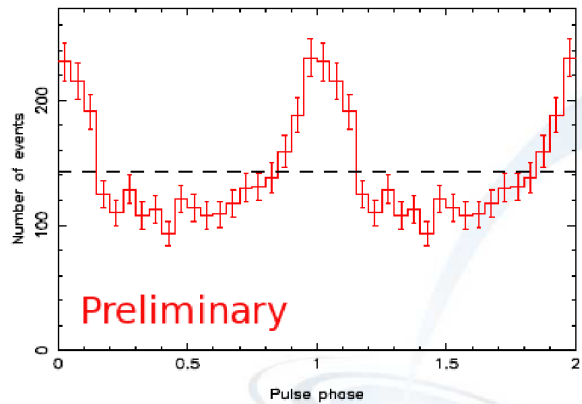}
  \caption{Pulse profile of PSR~J0734--1559, the latest $\gamma$-ray pulsar discovered in LAT blind searches.}
\label{fig1}
\end{figure}

{\bf A new Geminga-like pulsar in 1FGL J0734.7--1557} -- A recent blind search for pulsations using 2 years of data from the LAT unassociated source 
1FGL J0734.7--1557 has uncovered a new, middle-aged ($\tau\sim2\times10^5$ years) pulsar. PSR~J0734--1559 is a relatively faint ($\sim1\times10^{-7}$ ph 
cm$^{-2}$ s$^{-1}$) source located slightly off the Galactic plane (b$\sim2^\circ$). No previous X-ray coverage of this region exists and a short 
($\sim$5 ks) {\it Swift} observation did not reveal any likely counterparts. Deep follow-up radio observations at the Green Bank Telescope did 
not detect any radio pulsations from this pulsar. Figure~\ref{fig1} shows the single-peaked $\gamma$-ray pulse profile of PSR~J0734--1559.

\begin{figure}[htbp]
\includegraphics[height=.3\textheight]{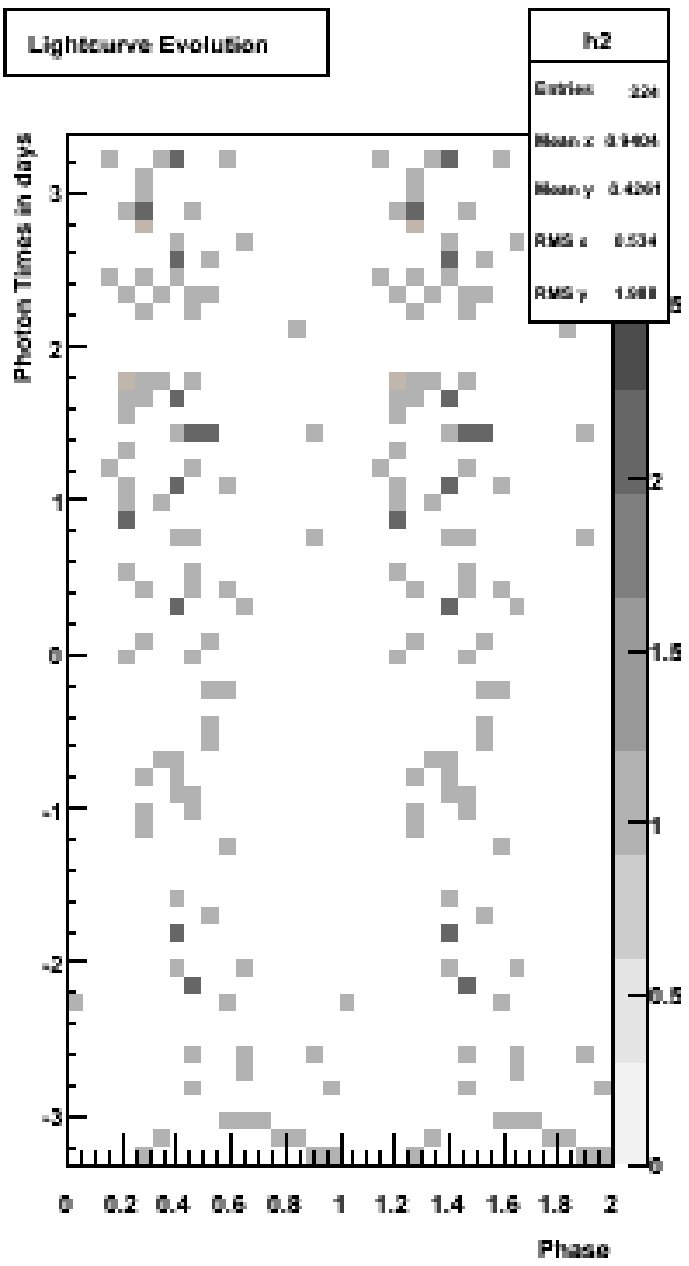}
\includegraphics[height=.3\textheight]{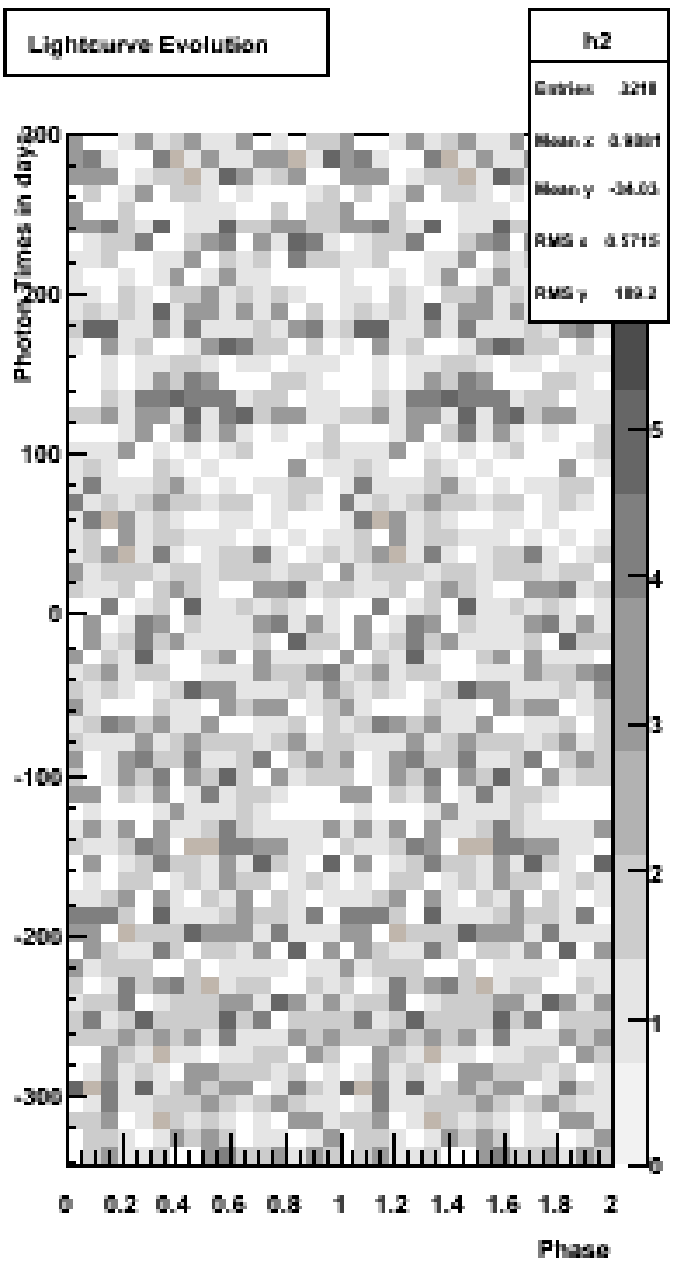}
\caption{Pulse profile evolution as a function of time for two pulsars found in blind searches of LAT data:
{\bf Left} -- CTA1, using 1 week of survey data ($\sim$100 photons). Search took 10 CPU minutes.
{\bf Right} -- PSR~J0734--1559, using 2 years of data ($\sim$1600 photons). Search took $\sim$200 CPU hours.}  
\label{fig2}
\end{figure}

{\bf Challenges --}
As the LAT continues to accumulate data and becomes sensitive to more and fainter $\gamma$-ray sources, blind searches for pulsations from 
these $\gamma$-ray sources become increasingly challenging and computationally expensive. Figure~\ref{fig2} illustrates this point. The left panel shows 
the result of a blind search on 1 week of survey data of the pulsar in CTA 1, the first pulsar discovered in blind searches of LAT data. A mere 
100 photons are sufficient for the blind search to ``discover" this pulsar at high significance, at the cost of only $\sim$10 CPU minutes. The panel 
on the right, however, shows the output from the search that uncovered the latest pulsar, PSR~J0734--1559. In this case, the search, using $\sim$2 years 
of survey data took over 1000 times longer to run. 

\begin{figure}
  \includegraphics[height=.3\textheight]{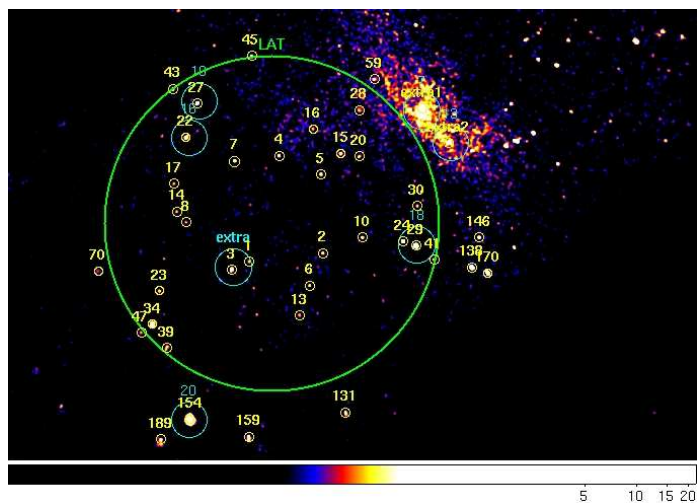}
  \caption{{\it Chandra} observation of the region around a LAT unassociated source (Credit: M. Marelli).}
\label{fig3}
\end{figure}

With these much longer integration times, two factors become increasingly important in improving the
sensitivity of our searches: an improvement in the search position and an improvement in the event selection. LAT positions, though significantly 
better than those of past $\gamma$-ray missions (e.g. EGRET) are still typically good only to several arc minutes. In particular, sources in the Galactic 
plane (where most young pulsars are expected to be located) are especially difficult to localize, due to systematic uncertainties in the diffuse 
emission model. One alternative to using LAT-determined positions is to use multiwavelength observations (typically X-rays) to identify a probable 
counterpart of a pulsar and barycenter the $\gamma$ rays at that position. Figure~\ref{fig3} shows a {\it Chandra} observation of a region around 
a LAT unassociated source, illustrating the much better angular resolution in X-rays, but also highlighting the challenge one faces in trying to 
identify, {\it a priori}, the potential X-ray counterpart of a LAT $\gamma$-ray source. A better event selection is also key to improving the 
sensitivity of blind searches. Most searches so far have relied on a simple {\it cookie cutter} selection of events around a particular source of 
interest. By performing a spectral analysis of a $\gamma$-ray source, however, it is possible to use the spectral model to improve the signal-to-noise 
by selecting events based on the likelihood that they come from our source of interest (see Belfiore, these proceedings).

{\bf Future Prospects} -- Blind searches of LAT data continue to yield results, as evidenced by the recent discovery of PSR~J0734--1559. However, the 
pace of discovery has slowed significantly, as expected. Future efforts will concentrate on improving the sensitivity of 
the current searches as described in the previous section, but also on expanding the parameter space to consider more exotic pulsars, such as radio-quiet 
millisecond pulsars. Finally, it is also interesting to consider the possibility of young radio-quiet $\gamma$-ray pulsars in binaries. The additional 
binary orbital parameters make a full blind search for such pulsars unfeasible, but a search for a pulsar in a binary of known orbital parameters 
(e.g. LS I+61 303) is a more realistic endeavor which is well worth pursuing.

\begin{theacknowledgments}
The {\it Fermi} LAT Collaboration acknowledges support from a number of agencies and institutes
for both development and the operation of the LAT as well as scientific data analysis. These include NASA and DOE in 
the United States, CEA/Irfu and IN2P3/CNRS in France, ASI and INFN in Italy, MEXT, KEK, and JAXA in Japan, and 
the K. A. Wallenberg Foundation, the Swedish Research Council and the National Space Board in
Sweden. Additional support from INAF in Italy and CNES in France for science analysis during the operations phase is also gratefully 
acknowledged.

\end{theacknowledgments}



\bibliographystyle{aipproc}   


\IfFileExists{\jobname.bbl}{}
 {\typeout{}
  \typeout{******************************************}
  \typeout{** Please run "bibtex \jobname" to optain}
  \typeout{** the bibliography and then re-run LaTeX}
  \typeout{** twice to fix the references!}
  \typeout{******************************************}
  \typeout{}
 }

\def \apjl {ApJL}
\def \apj {ApJ}

\bibliography{ms}

\end{document}